\documentclass[aps,twocolumn,showlabels,showrefs,amsmath,amssymb,prr,superscriptaddress,floatfix,colors]{revtex4-1}

\usepackage{graphicx}
\usepackage{dcolumn}
\usepackage{bm}
\usepackage{amssymb}
\usepackage{multirow}
\usepackage{color}
\usepackage{braket}
\usepackage[normalem]{ulem}
\usepackage{mathtools}

\usepackage{xcolor}
\definecolor{blueprl}{RGB}{13.0, 18.0, 180.0 }
\usepackage[colorlinks=true, allcolors=blueprl]{hyperref}
\definecolor{massimiliano}{RGB}{0,0,255}

\def\ri{\bm{r}_i}
\def\rj{\bm{r}_j}
\def\dri{\dot{\textbf{r}}_i}
\def\ddri{\ddot{\textbf{r}}_i}

\begin{document}

\title{Arrested phase separation and chiral symmetry breaking in active dumbbells under shear}

\author{Lucio Mauro Carenza}
\affiliation{Dipartimento  Interateneo di  Fisica,  Universit\`a  degli  Studi  di  Bari  and  INFN,
Sezione  di  Bari,  via  Amendola  173,  Bari,  I-70126,  Italy}

\author{Giuseppe Negro}
\affiliation{SUPA School of Physics and Astronomy, University of Edinburgh, Peter Guthrie Tait Road, Edinburgh, EH9 3FD, UK}

\author{Pasquale Digregorio}
\affiliation{Dipartimento Interateneo  di  Fisica,  Universit\`a  degli  Studi  di  Bari  and  INFN,
Sezione  di  Bari,  via  Amendola  173,  Bari,  I-70126,  Italy}

\author{Antonio Suma}
\affiliation{Dipartimento Interateneo  di  Fisica,  Universit\`a  degli  Studi  di  Bari  and  INFN,
Sezione  di  Bari,  via  Amendola  173,  Bari,  I-70126,  Italy}

\author{Giuseppe Gonnella}
\affiliation{Dipartimento  Interateneo di  Fisica,  Universit\`a  degli  Studi  di  Bari  and  INFN,
Sezione  di  Bari,  via  Amendola  173,  Bari,  I-70126,  Italy}

\email[email: ]{name@}

\begin{abstract}
  Through molecular dynamics simulations, we investigate the phase separation and aggregation dynamics of active dumbbell particles in two-dimensions subjected to shear. 
  We find that the growth of the phase-separated region is arrested when shear is applied, with the average clusters size plateauing towards a value $R_s$ that remains constant over time. While activity enhances the resilience of clusters against shear-induced breakup,  $R_s$  decreases with growing shear rate $\dot\gamma$, with an intermediate regime where $R_s\propto \dot\gamma^{-1}$. We find that clusters in the stationary state are progressively less polarized and  increasingly elongated with increasing shear. At the same time, we find a breaking in chiral symmetry of both rotation direction and internal organization of clusters: typically, dumbbells point towards the cluster center with a small non-zero angle, such that the active torque opposes the shear torque, with cluster's angular velocity well captured by a simplified analytical model.  We argue this conformation makes clusters more stable against shear.   
\end{abstract}

\maketitle

Systems undergoing phase separation have a special place in non-equilibrium physics~\cite{bray1994,Onuki_2002,yeomans2009,Cates_Tjhung_2018}, showing very general properties, such as dynamical scaling, universal growth exponents, etc.
The dynamical behavior can be strongly affected by the presence of external driving~\cite{Gonnella1998,Bray2001,saracco2005,cates2006,Gonnella2007pre,Cates_Tjhung_2018,Biroli2024,Allen2008}. This 
 generally counteracts the effects of thermodynamic forces, which tend to drive the system toward free-energy equilibrium.  The interplay between these opposing influences can lead to different dynamics.
For instance, in phase-separating diffusive binary mixtures under shear flow, domain growth occurs indefinitely but at different rates depending on the direction, with domains elongating along the flow~\cite{Gonnella1998,Onuki1995,rapapa1999}. Conversely, in binary fluids, a non-equilibrium steady state emerges, where the characteristic length scale of the domain structure  attains a finite value at late times~\cite{Wagner1999,cates2006}. 
Besides these well established results, other relevant questions remain  regarding how morphological and dynamical properties change under external driving in other important classes of phase-separating systems, especially those that are inherently out of thermodynamic equilibrium, such as active systems.

Active matter presents  rich complex dynamical behaviors~\cite{ marchetti2013, elgeti2015, bechinger2016, fodor2018, shaebani2020,Head2024,caprini2021,CaporusooNegro2024,Caprini2023,Semeraro_2024,Alert2024}, in particular when the effects of  external driving combine with those of self-propulsion.
For instance,  active fluids show non-Newtonian behaviors ~\cite{marchetti2013,Santillan2018,Koch2011,Bayram2023, Mao2024,lamuramath2021,Mandal2021,Goswami2025,Roca2025},
or even  a negative effective viscosity~\cite{Lopez2015,Favuzzi2021}. 
A fundamental property of self-propelled particles is their ability to spontaneously aggregate despite the absence of attractive interactions, triggering   
 a motility-induced phase separation (MIPS) between a low-density gas-like phase and dense stable aggregates \cite{catesMIPS,fily2012,Winkler2016,gonnellaMIPS,rednerMIPS,buttSPP,suma2014motility,digregorio2017, caporusso2020motility, caporusso2023,speckCRYSTALSPP,bechinger2016,Omar2020}, reminiscent of the equilibrium liquid-gas transition but in the absence of cohesive forces.  One thus can ask what is the effects of external driving on the kinetics of MIPS, and we will do so considering an important  class of active polar particles systems.

Indeed, many active matter systems are composed of intrinsically polar units. Examples are motile cells~\cite{Picamal2012}, bacteria~\cite{Dombrowski2004}, or synthetic self-propelled rods~\cite{Paxton2004}, each of which has a well-defined head–tail direction. This intrinsic polarity underlies their ability to move persistently in a preferred direction, which not only shapes their individual trajectories but also orchestrates large-scale collective behaviors~\cite{marchetti2013,chateDADAM,Toner1998,Paoluzzi2024,Athani2024,WEBER2013,Geyer2019}.
Here we focus on a paradigmatic model of self propelled polar units, i.e.  active Brownian dumbbells~\cite{
Linek2012,suma2014dynamics} (ABD), composed of particles having an anisotropic shape formed by two connected beads, with an active force applied along their axis. ABD on one hand exhibit MIPS~\cite{suma2014motility,digregorio2017} with a substantially faster domain growth than active disks~\cite{caporusso2024phase,gonnella2014phase,digregorio20192d}, while on the other hand coalesce in clusters with strong polar properties and hexatic order~\cite{suma2014motility,digregorio2017,Petrelli2018}, that give rise to  sustained rotational and translational motions~\cite{caporusso2024phase}. Consequently, ABD offer a valuable model to investigate whether shear can (i) influence cluster growth and (ii) affect the clusters’ internal polarization properties, thus shedding light on the effect of external driving on the self organized structural order of active systems.

To address these questions, we  study the phase-separation kinetics of ABD under shear in two dimensions, using molecular dynamics simulations, in the  MIPS region established in Refs.~\cite{gonnella2014phase,suma2014motility,digregorio2018,Petrelli2018}.
By examining the time evolution of the average domain size  $R(t)$, we show that under shear the  phase separation process is arrested, plateauing toward a value $R_s$, which depends both on activity and magnitude of external driving.
Next, we examine the internal organization of stationary clusters under shear.  First, the magnitude of the local polarization decreases with increasing shear. 
At the same time, the torque imparted by active forces is non-zero and opposes that of the shear, leading to a preferred dumbbell orientation pattern that breaks cluster chirality. We show that clusters with an active torque opposing the shear are indeed more stable than those with an active torque acting in the same direction.

{\bf Model} - 
We consider $N$ self-propelled dumbbells, composed by two beads, each of size $\sigma$ and mass $m$. Each dumbbell's bead $i=1,...2N$ moves according to the following Langevin equation \cite{suma2014motility,suma2014dynamics}:

\begin{equation}
\begin{split}
    m\ddri&=-\Gamma \dri+\bm{F}^{\bm{a}}_i-\sum_{i\neq j}\nabla U_i(r_{ij})\\ &+\sqrt{2\Gamma k_B T}~\bm{\xi}_i+\Gamma\dot\gamma y_i \bm{e_x}, 
\label{eq:lang_eqs}
\end{split}
\end{equation}
where $\ri=(x_i,y_i)$ is the position of the $i$-th bead, the dots denote the time derivatives, $r_{ij}=|\ri-\rj|$ is the distance between the $i$-th and the $j$-th bead, $\Gamma$ is the friction coefficient, $k_B$ the Boltzmann constant, $T$ is the bath temperature and $\bm{\xi}_i$ is an independent zero-mean white noise which satisfies $\braket{\bm{\xi}_i(t)\bm{\xi}_j(s)}=\delta_{ij}\delta(t-s)\bm{1}$. 
$\bm{F}^{\bm{a}}_i$ is an active force, applied to each bead, that self-propels the dumbbell, with constant modulus $F^a$ and pointing along the tail-to-head direction of each dumbbell.
The force term $\Gamma\dot\gamma y \bm{e_x}$ accounts for the shear, with $\dot\gamma$ the shear rate and $\bm{e_x}$ the x-axis unit vector. 

$U$ is the potential between any pair of beads, comprised of 
a FENE (finitely extensible nonlinear elastic) potential acting between two beads of the same dumbbell, $U_{FENE}(r)=-0.5kr_0^2\ln[1-(r/r_0)^2]$, with $r_0$ the maximum extent of the bond and $k$ the elastic constant, and a Weeks-Chandler-Anderson (WCA) potential---a purely repulsive truncated and shifted Lennard-Jones (LJ) potential---acting between any pair.
$U_{WCA}(r)=4\epsilon [(\sigma/r)^{12}-(\sigma/r)^{6}]+\epsilon$ if $r<r_c=2^{1/6}\sigma$ and $0$ otherwise, with $\epsilon$ the energy unit when two spheres are separated by a distance $\sigma$ and $r_c$ the location of the potential minimum. 
We fix the number of dumbbells to $N=512^2$, and vary the system size $L$ such that the packing fraction is $\phi=2N\pi\sigma^2/4L^2\sim 0.4$ (we will consider a different density $\phi=0.5$ only in Fig.~\ref{comparison_densities}), and we employ Lees-Edwards periodic boundary conditions~\cite{Lees_1972}. 
A relevant quantity for the system is the Péclet number $\textrm{Pe}=2 F^a\sigma/k_BT$~\cite{digregorio2017}, which compares the strength of activity and thermal fluctuations. Another relevant parameter is the shear Péclet number $\textrm{Pe}_s=\dot\gamma\sigma^2\Gamma/k_BT$, which quantifies the relative influence of the imposed shear flow compared to thermal fluctuations. 
All quantities are expressed in units of mass $m$, length $\sigma$, and energy $\epsilon$, which are set to one. 
The simulation time unit, $\tau_{LJ}=\sqrt{m\sigma^2/\epsilon}$, is also unitary, and the timestep is set to $\Delta t=0.01$.
We also set $\Gamma=10$,  $k_BT=0.01$, $r_0=1.5$ and $k=30$. We perform simulations with $\textrm{Pe}=100, 150, 200, 400$ (corresponding to $F^a=0.5, 0.75, 1.0, 2.0$), 
and shear rates in the range $\textrm{Pe}_s=[0:5]$ (corresponding to $\dot\gamma=[0:5\times 10^{-3}]$).
The Pe values are chosen such that the system at the given density phase separates through MIPS at $\textrm{Pe}_s=0$~\cite{gonnella2014phase,suma2014motility}. 
For each set of parameters, we perform $20$ independent simulations.
We use LAMMPS~\cite{plim1995} to integrate numerically the equations of motion, with a custom code for the implementation of dumbbells self-propulsion. 
Simulations are typically run for about $5\times10^{7}$ time units.

\begin{figure}[t!]
\centering
       \includegraphics[width=1.0\columnwidth]{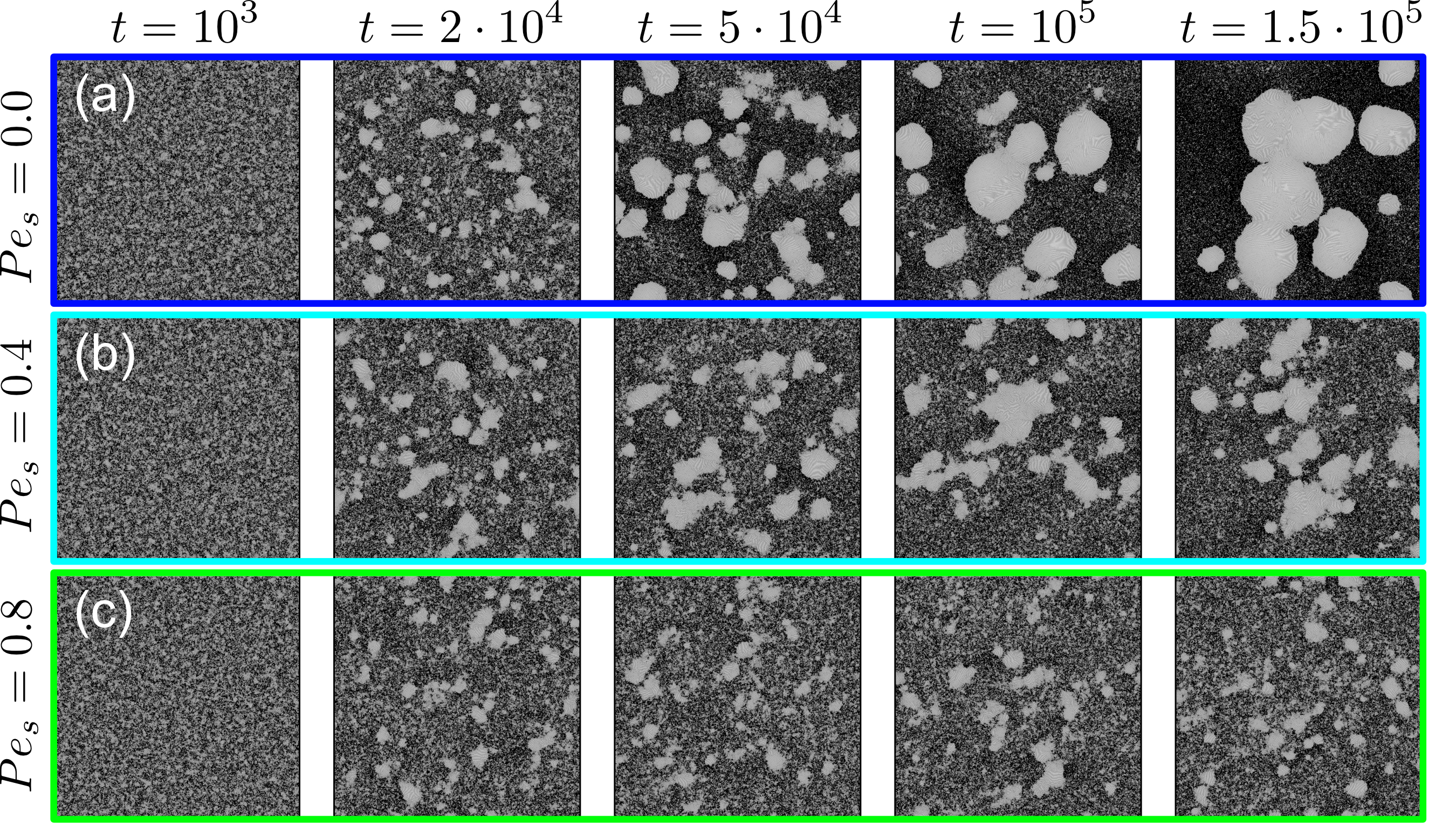}
\caption{\textbf{The phase separation process.} Snapshots of the systems at Pe=200, $\phi=0.4$ and (a) $\textrm{Pe}_s=0$, (b) $\textrm{Pe}_s=0.4$, and (c) $\textrm{Pe}_s=0.8$,  at  different times. 
For $\textrm{Pe}_s=0$, the first two snapshots correspond to the nucleation regime, the third is taken in the intermediate time evaporation-condensation regime, while in the last two snapshots clusters grow by collisions and merging. For $\textrm{Pe}_s=0.4$ and $\textrm{Pe}_s=0.8$ growth is arrested after the third and second panel, respectively.}
\label{Fig1}
\end{figure}

{\bf Growth kinetics under shear} - We start by considering the case without applied shear, $\textrm{Pe}_s=0$ and $\mathrm{Pe}=200$.
Fig.~\ref{Fig1}(a) shows snapshots of the time evolution of the dumbbells system starting from a random configuration. 
We find here a first regime where nucleation of clusters begins, for $t\lesssim 10^4$, followed by a time interval where clusters grow in size through condensation of dumbbells around the nucleated clusters. At late times, clusters are observed to grow through translation, collision and merging.
This behavior is qualitatively similar to the one found in~\cite{caporusso2024phase,noterigid}, although in that case rigid dumbbells were considered.

We then turn on the shear, considering  $\textrm{Pe}_s=0.4$ and $Pe=200$.  Fig.~\ref{Fig1}(b), middle row, shows snapshots of the time evolution, starting from the same random configuration and taken at the same time as in Fig.~\ref{Fig1}(a). Here, we observe at a similar timescale the nucleation of clusters (first two panels), followed by their growth (third panel). However, in the late regime (last two panels) clusters arrest their growth. The same happens for larger shear, $\textrm{Pe}_s=0.8$ and $Pe=200$, shown in Fig.~\ref{Fig1}(c) (bottom row). In this case, growth arrests sooner with clusters reaching  a smaller stationary  size.

\begin{figure}[t!]
\centering
       \includegraphics[width=1.0\columnwidth]{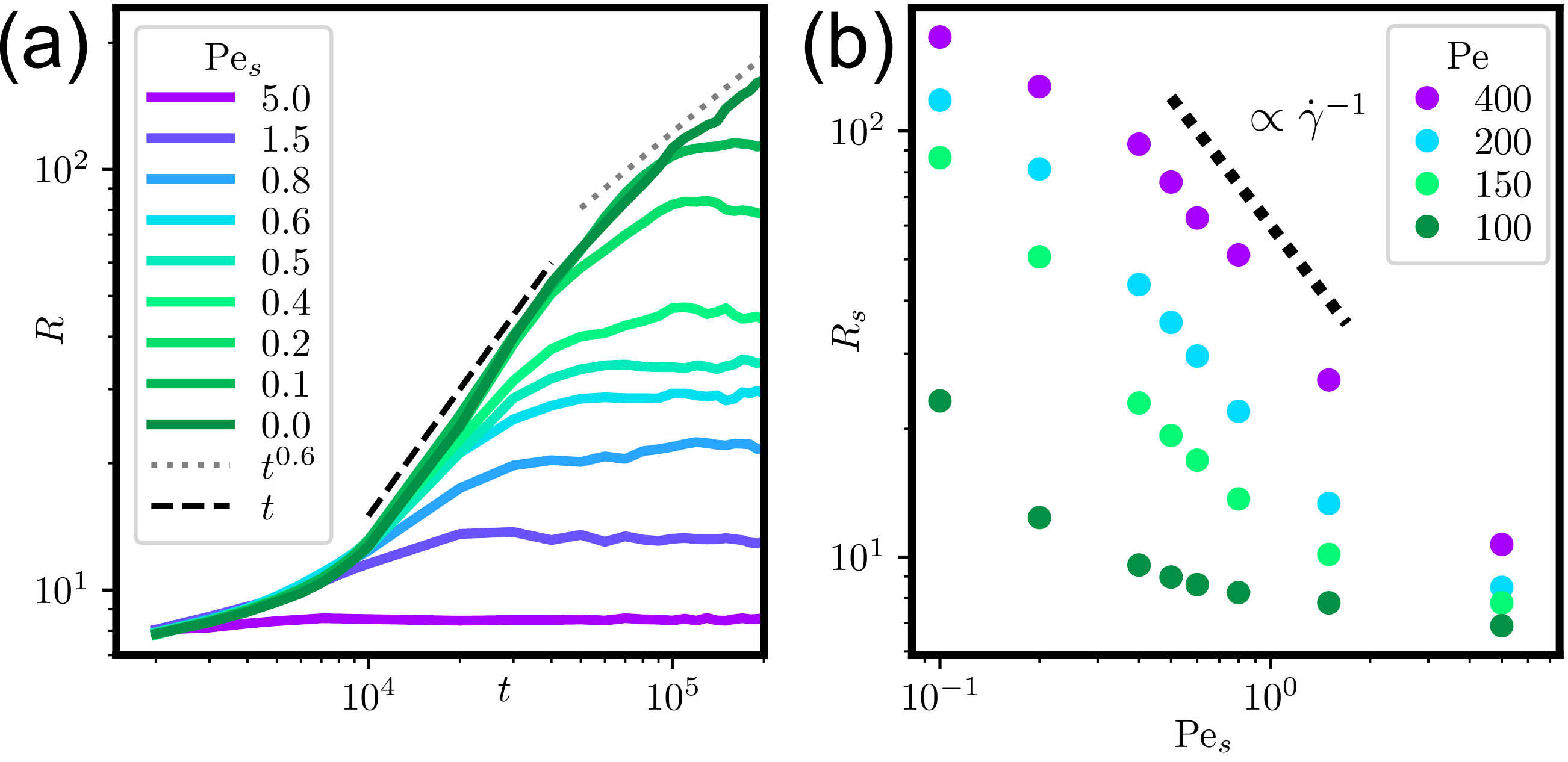}
\caption{\textbf{Growth kinetics under shear.} (a) Growth length $R(t)$  as a function of time for  $Pe = 200$ and  global packing fractions $\phi = 0.4$, for different  $\textrm{Pe}_s$. Each curve has been averaged over $20$ independent runs. The dashed and  dotted lines represent $t$ and $t^{0.6}$, respectively. 
(b) Average value of cluster size $R_s$ as a function of shear rate, computed in the stationary long time regime, for $Pe=100, 150, 200, 400$ and $\phi=0.4$. As dashed line, slope proportional to $1/\dot\gamma$.}
\label{Fig2}
\end{figure}

In order to characterize quantitatively the clusters growth, we measure the
typical length of the dense component, $R(t)$, as the inverse average over the structure factor, $S(\mathbf{k},t)$, of the modulus $k$ of the wave vector in Fourier space~\cite{gonnella2014phase,caporusso2020motility,caporusso2024phase} 

$R(t)=\pi\int d\mathbf{k} S(\mathbf{k},t)/\int d\mathbf{k} k S(\mathbf{k},t)$, with the structure factor defined as
$S(\mathbf{k},t)=\frac{1}{2N}\sum_{i}^{2N}\sum_{j}^{2N}e^{i\mathbf{k}\cdot(\mathbf{r}_i(t)-\mathbf{r}_j(t))}$.

The growth length $R(t)$ is plotted in Fig.~\ref{Fig2}(a) for Pe=200 and various values of $\textrm{Pe}_s$ (see Fig.~\ref{AppFig1} for Pe=100, 150). Starting from no shear ($\textrm{Pe}_s=0$), we can identify the  three regimes observed in Fig.~\ref{Fig1}(a); a nucleation period below $t\lesssim 10^4$, followed by a rapid growth regime until $t\lesssim 5\times 10^4$, with $R(t)\sim t$ (see dashed line). The last regime, instead, corresponding to the clusters growth through collisions, scales with $R(t)\sim t^{0.6}$, compatible with the exponent found for rigid dumbbells~\cite{caporusso2024phase}.

When shear is applied with $\textrm{Pe}_s\lesssim 1.5$, the first nucleation regime is always present, and curves overlap with each other. For $t \geq 10^4$, however, the subsequent evolution depends on the shear rate.
With small enough shear applied, $\textrm{Pe}_s\leq 0.2$, we are still able to see the rapid growth regime, while the late regime is less pronounced, as $R(t)$ tends toward a constant value.
For larger shear, $0.2 \leq\textrm{Pe}_s\lesssim 1.5$, the rapid growth regime is not visible anymore, with curves plateauing for $t\gtrsim 3\times 10^4$.  Finally, for large enough shear, $\textrm{Pe}_s> 1.5$, the nucleation regime completely disappears and curves immediately plateau. In general, we always observe stationary values of $R(t)$ at large times, $R_s$, that decrease with increasing shear.

Fig.~\ref{Fig2}(b) shows values of $R_s$ for all Pe and $\textrm{Pe}_s$ considered at $\phi=0.4$ (see also Fig. \ref{comparison_densities} for $\phi=0.5$). Two important characteristics emerge: $R_s$ is a  decreasing function of shear, while it increases with activity. 
Thus, it becomes evident that while shear plays a role in breaking up clusters, activity plays the opposite role in stabilizing clusters. Moreover, we observe in these curves an intermediate regime of Pe$_s$ where $R_s$ scales as $R_s \sim 1/\dot\gamma $.

To understand the observed scaling and the arrest of cluster growth, we propose a simple force-balance argument for dumbbells located at the cluster periphery. Activity generates an inward cohesive force, consistent with the mechanism underlying MIPS. On the other hand, the imposed shear flow exerts a drag that preferentially strips particles from the side of the cluster exposed to the flow. When the shear-induced drag becomes comparable to the active cohesive force, a balance emerges between the influx of particles and those escaping, resulting in a finite steady-state cluster size. The observed scaling of $R_s$ is consistent with the above argument, since the shear-induced force acting on the cluster surface relative to its center of mass scales as $\Gamma \dot\gamma R_s$, and balancing this with the active force yields $R_s \propto 1/\dot\gamma$. We stress, however, that this argument neglects other relevant contributions to cluster breakdown. These include cluster distortion under shear, which promotes internal structural rearrangements, as well as defects and grain boundaries, which can act as weak points for fragmentation~\cite{digregorio2022}. At sufficiently large shear rates, $\textrm{Pe}_s> 1.5$, the nucleation regime completely disappears, and thus the effect of MIPS becomes negligible; in this case, we expect cluster sizes to be controlled only by fluctuations driven by both shear and activity.

We next examine how shear affects the internal arrangement of dumbbells within clusters, and how it modifies their dynamics.

\begin{figure}[t]
\centering
       \includegraphics[width=1.0\columnwidth]{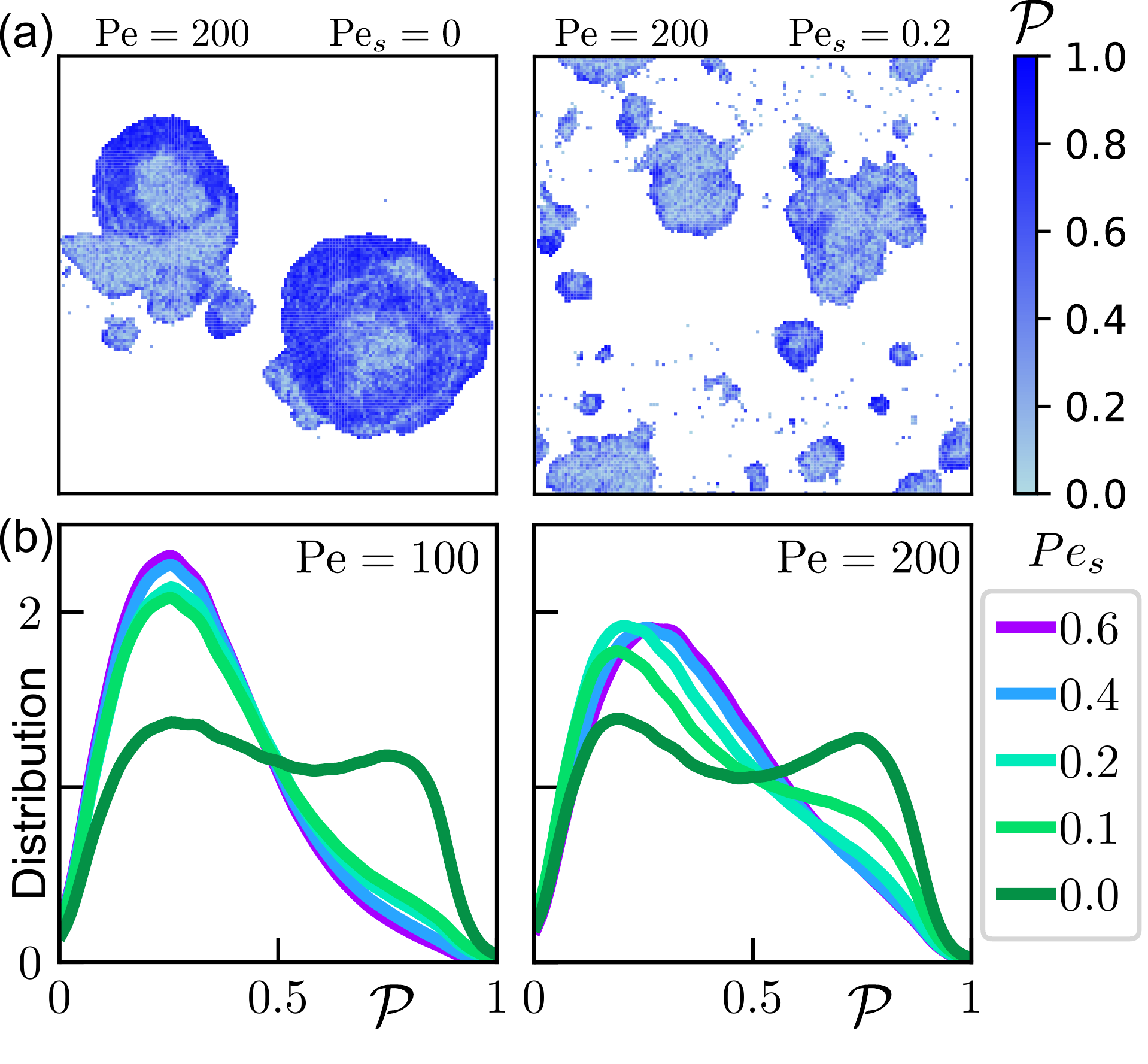}
\caption{\textbf{Local polarization.} (a) Snapshots of the system at $Pe=200$, and $\textrm{Pe}_s=0.0, 0.2$, with dumbbells in clusters colored according to the magnitude of the local polarization $\mathcal{P}$.   (b) Probability distribution of $\mathcal{P}$ inside clusters for Pe $= 100, 200$ (left and right panel) for different values of  $\textrm{Pe}_s$.
}
\label{Fig3}
\end{figure}

\begin{figure}
       \includegraphics[width=1.0\columnwidth]{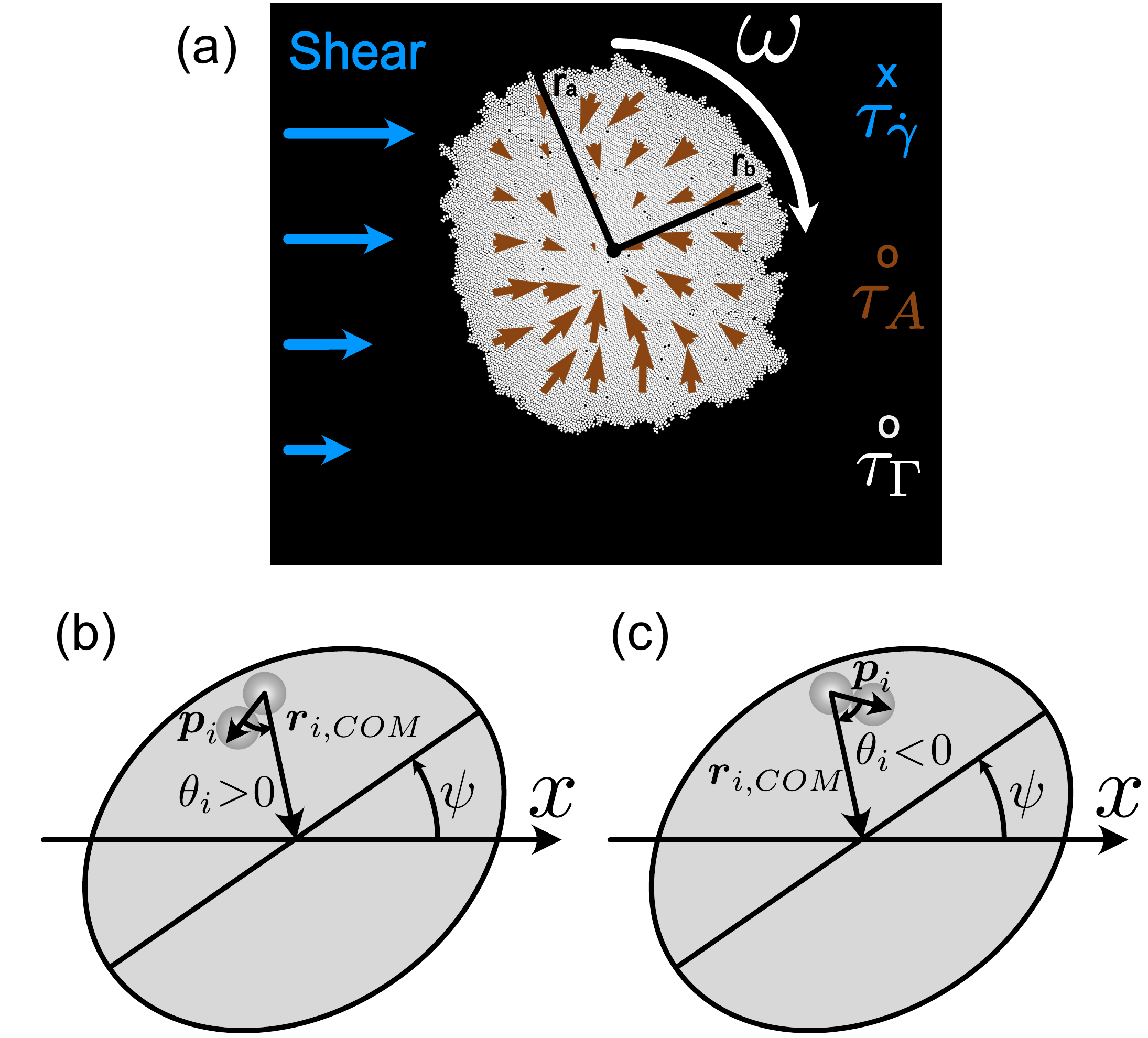}
       \caption{\textbf{Cluster properties.} (a) Schematic organization of a cluster. Brown arrows indicate the polarization, forming a spiral pattern,  $r_a$ and $r_b$ the major and minor semi-axis of the cluster computed from the inertia tensor. Blue arrows indicate the direction and magnitude of applied shear. Highlighted are also the average directions of various torques ( $\circ$,   $\times$ representing the outward (outgoing) and inward (incoming) directions), and of the angular velocity $\omega$ of the cluster. Panels (b) and (c) give a schematic representation of the definition of  $\theta_i$, the signed angle between $\bm{p}_i$, the dumbbell tail-to-head versor,  and the vector connecting the dumbbell and the cluster center of mass.
       }
\label{Figscheme}
\end{figure}

\begin{figure}
       \includegraphics[width=1.0\columnwidth]{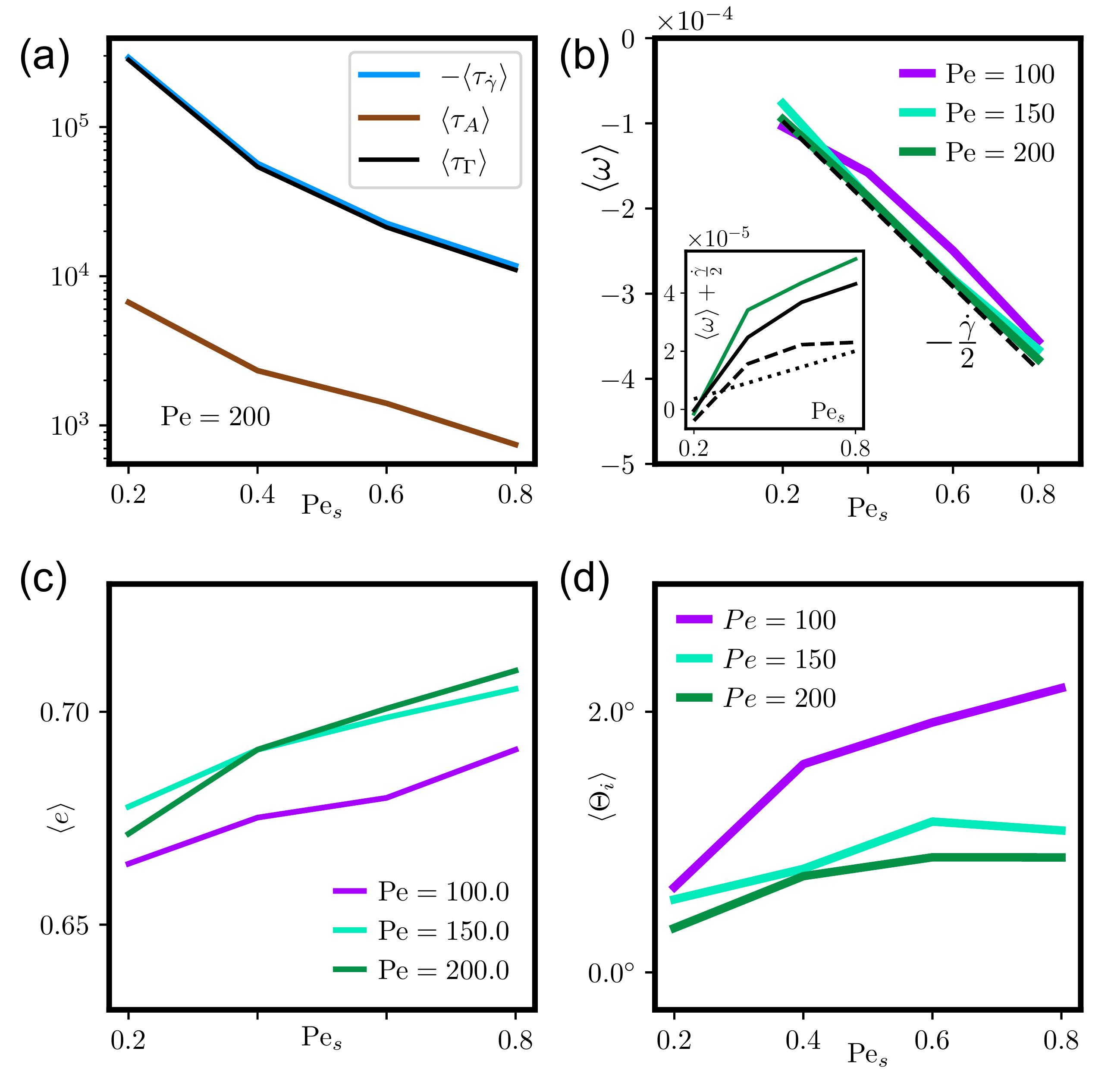}
       \caption{\textbf{Cluster properties.} (a)  Torques contributions for Pe=200. (b) $\langle\omega\rangle$ as a function of $\textrm{Pe}_s$ at different Pe. The dashed line represents the linear dominant dependence $-\dot\gamma/2$. In the inset, $\langle\omega\rangle+\dot\gamma/2$ for Pe=200 (green line). For comparison, the black continuous curve shows the predicted values $\hat\omega+\dot\gamma/2$},  Eq.~\ref{omegahat}, given by the sum of  $\frac{4}{3}\frac{F^a \sin{\theta}}{\Gamma r}$ (dashed line),  and a term proportional to $e^4$   (dotted line). $\langle\omega\rangle$ and values of $r$ and $\theta$ imputed in Eq.~\ref{omegahat} are obtained averaging over all clusters. (c) Average clusters eccentricity, at various Pe and $\textrm{Pe}_s$. (d) Average dumbbell orientation $\langle\theta_i\rangle$. 
\label{Fig4}
\end{figure}

{\bf Structural and dynamical properties} - One notable feature of self-propelled dumbbells, in absence of shear, is the formation of rotating clusters that exhibit a strong polarization, and are able to rotate at a constant angular velocity due the balance between the active and drag torques ~\cite{suma2014motility,Petrelli2018,caporusso2024phase}. The active torque, in particular, originates from dumbbells which are oriented at a certain angle with the cluster center of mass, forming a polar field with a spiral pattern~\cite{suma2014motility,Petrelli2018}. These observations invite the question of how such self-sustained dynamics and emergent polar order respond to external driving. 

To explore this, we first analyze the internal organization of clusters under shear, in the stationary regime reached at late times. Clusters are identified using 
 the DBSCAN algorithm\cite{dbscan}, with minimum number of neighbors $3$ and cutoff distance $1.1$. 
 For each cluster we compute the local polarization field $\bm{\mathcal{P}}$~\cite{Petrelli2018} by first identifying, for each dumbbell in a cluster, the tail-to-head versor, $\bm{p}_i$, and then averaging it over bins of size $5\sigma$. We consider only bins whose local area fraction $\phi$ exceeds $\phi_c=0.6$, and thus pertain to a cluster. Note that without shear one expects clusters to be highly polarized for the values of Pe considered~\cite{Petrelli2018}.

Fig.~\ref{Fig3}(a) compares two snapshots of the norm of the local polarization field, $\mathcal{P}$, for $\textrm{Pe}_s=0, 0.2$ at Pe=200. We note that, whether shear is present or not, the local polarization is largest near the edges of the clusters; however, in the absence of shear, the region of high polarization is more extended. Previous works in the absence of shear have shown that this arrangement is strongly influenced by several factors~\cite{Petrelli2018}, including density, Péclet number, initial configuration, and system size, as cluster merging can substantially alter the internal arrangement of dumbbells.
These observations regarding the local polarization are supported by the probability distribution functions of $\mathcal{P}$ in Fig.~\ref{Fig3}(b). At both $\textrm{Pe}=100$ and $\textrm{Pe}=200$ with $\textrm{Pe}_s=0$, the distribution exhibits two peaks: one associated with strongly polarized dumbbells near the cluster boundary, and another corresponding to more weakly polarized dumbbells in the cluster interior.
When shear is introduced, the second peak in the distribution is gradually suppressed, reflecting a reduction in the degree of dumbbells alignment near the cluster boundary. This effect parallels the overall decrease in cluster size induced by the shear flow. Moreover, the attenuation of the second peak appears to depend on the active force. For instance, at $\textrm{Pe}=100$, even moderate shear significantly lowers the second peak, while at $\textrm{Pe}=200$, a noticeable second peak persists at moderate shear values, before ultimately diminishing at higher shear rates. 

We can motivate this decrease in polarization by accounting for the fact that the peripheral dumbbells—which are primarily responsible for higher polarization—are the most affected by shear. Shear acts directly on the cluster surface, disrupting alignment; at the same time, the reduction in cluster size hinders the formation of well-organized peripheral regions. Moreover, shear also distorts the cluster globally, altering the internal arrangement of dumbbells and thereby affecting the local polarization.

We now turn to characterize the dynamics of clusters. In the absence of shear, a balance between the total active and drag torques acting on the cluster, ${\bm\tau}_A+{\bm\tau}_\Gamma=0$, determines the rotating velocity $\omega$ of clusters~\cite{caporusso2024phase}, with ${\bm\tau}_A=\sum_{i=1}^{N_c} \bm{r'}_i \times\bm{F}^{\bm{a}}_i$ and ${\bm\tau}_\Gamma=\sum_{i=1}^{N_c}\bm{r'}_i \times (-\Gamma{\dot{\bm{r}}}'_i) $,  $\bm{r'}_i = \bm{r}_i - {\bm{R}_{\text{COM}}}$  the position of the $i$-th bead with respect to the cluster's 
center of mass, and $N_c$  the total number of beads in a cluster.
With shear, an additional torque contribution modifies such balance, ${\bm\tau}_A+{\bm\tau}_{\dot\gamma}+{\bm\tau}_\Gamma=0$, with ${\bm\tau}_{\dot\gamma}=\sum_{i=1}^{N_c}\bm{r'}_i \times \Gamma\dot\gamma y_i \bm{e_x}$ (Fig.~\ref{Figscheme} (a)). 

We independently compute the three torque contributions and averaged them over all clusters in the stationary regime, focusing on intermediate $\textrm{Pe}_s$ values where cluster nucleation through MIPS occurs. Results are shown in Fig.~\ref{Fig4}(a) for Pe=200 (see Appendix B and Fig.~\ref{AppFigtorque} for other values of Pe). 
First, we verify that the sum of the three terms balance out, with ${\tau}_{\dot\gamma}$ and ${\tau}_A$ opposite in sign. In particular, shear applies a negative (clockwise) torque, while the active force applies a much smaller positive (counter-clockwise) torque. The difference between these two terms, balanced by drag, determines a clockwise rotation, providing a breaking in the rotational symmetry with respect to the case $Pe_s=0$ (see also Appendix B and C for other cases).
Indeed, measuring the average angular velocity of clusters $\langle\omega\rangle$, we find it is always negative on average, linearly decreasing with $\textrm{Pe}_s$, and slightly affected by Pe, Fig.~\ref{Fig4}(b), further pointing to the major role played by shear in determining the value of $\langle\omega\rangle$.

This behavior can be rationalized  with a simple model.  Since clusters are substantially distorted by shear, as can be appreciated looking at the two semi-axes, $r_a$ and $r_b$ of each cluster (Fig.~\ref{Figscheme}(a)), we compute the eccentricity $e=\sqrt{1-\frac{r_b^2}{r_a^2}}$, displayed in Fig.~\ref{Fig4}(c).
We approximate a cluster of dumbbells as a  homogeneous ellipse,  with each volume element subjected to shear, and to an active force with an orientation angle  $\theta$  with respect to the center.  By modeling the ellipse as rotating at  angular velocity $\omega$ with respect to its center, and tilted with a certain angle $\psi$ with respect to the x-axis, we compute the total torques acting on the ellipse ${\tau}_{\dot\gamma}=-\dot\gamma\Gamma[\frac{\pi r_ar_b}{4}(r_a^2\sin^2\psi+r_b^2\cos^2\psi)]$,  ${\tau}_A=\mathrm{F}_a\sin(\theta)\frac{4r_a^2r_b}{3}E(e)$, $\tau_\Gamma=-\Gamma\omega(\frac{\pi r_ar_b}{4})(r_a^2+r_b^2)$ with $E$ an elliptic integral (see Appendix C for further discussion). Balancing the torques, one obtains a first order differential equation in $\psi$ (with $\omega=d\psi/dt$). For typical parameters of our system, the equation admits  periodic solutions of $\omega$ (see Appendix C for further discussion), with a time average $\hat\omega$ that can be expanded for small eccentricity $e$ as  (see Ref.~\cite{jeffery1922motion} for the passive case):
\begin{equation}
    \hat\omega=\omega _{0}+e^{4}\left( -\dfrac{5}{48}\dfrac{F^a \sin \theta }{\Gamma r}-\dfrac{\dot\gamma^2}{32}\dfrac{1}{\omega _{0}}\right)+O(e^6)
\label{omegahat}
\end{equation}
with $\omega _{0}=\frac{4}{3}\frac{F^a \sin{\theta}}{\Gamma r} - \frac{\dot\gamma}{2}$ the value of the angular velocity for a circle of radius $r$ (in the case $r_a=r_b=r$).
Despite the simplicity of the model, we find good agreement when  comparing measured and predicted torques values (see Appendix C and Fig.~\ref{Figpredizione}). In Eq.~\ref{omegahat} we see that $\hat\omega$ has a dominant linear dependence on $\frac{\dot\gamma}{2}$, also observed in simulations (dashed line in Fig.~\ref{Fig4}(b)), an   additional dependency proportional to activity, $\frac{4}{3}\frac{F^a \sin{\theta}}{\Gamma r}$, and a first correction term due to the elliptical shape. The term $\frac{\dot\gamma}{2}$ is an order of magnitude larger than the other two, which are depicted as dashed and dotted lines in the inset of Fig.~\ref{Fig4}(c),  alongside their sum (black line), to be compared with the one measured in simulations $\langle\omega\rangle$ (green line).

To assess the actual internal dumbbells organization inside clusters, we compute the average orientation angle, $\langle\theta_i\rangle$(Fig.~\ref{Fig4}(d)), with $\theta_i$ the signed angle between $\bm{p}_i$ and the vector connecting the dumbbell and the cluster center of masses (see Fig.\ref{Figscheme}(b)-(c) for a representation).
We find this angle to be positive and non-zero, meaning that dumbbells are indeed oriented on average in such a way to produce a positive active torque that opposes the one imparted by shear, in line to the other measurements (Fig. \ref{Fig4}(a)).

\begin{figure}[t!]
\centering
       \includegraphics[width=1.0\columnwidth]{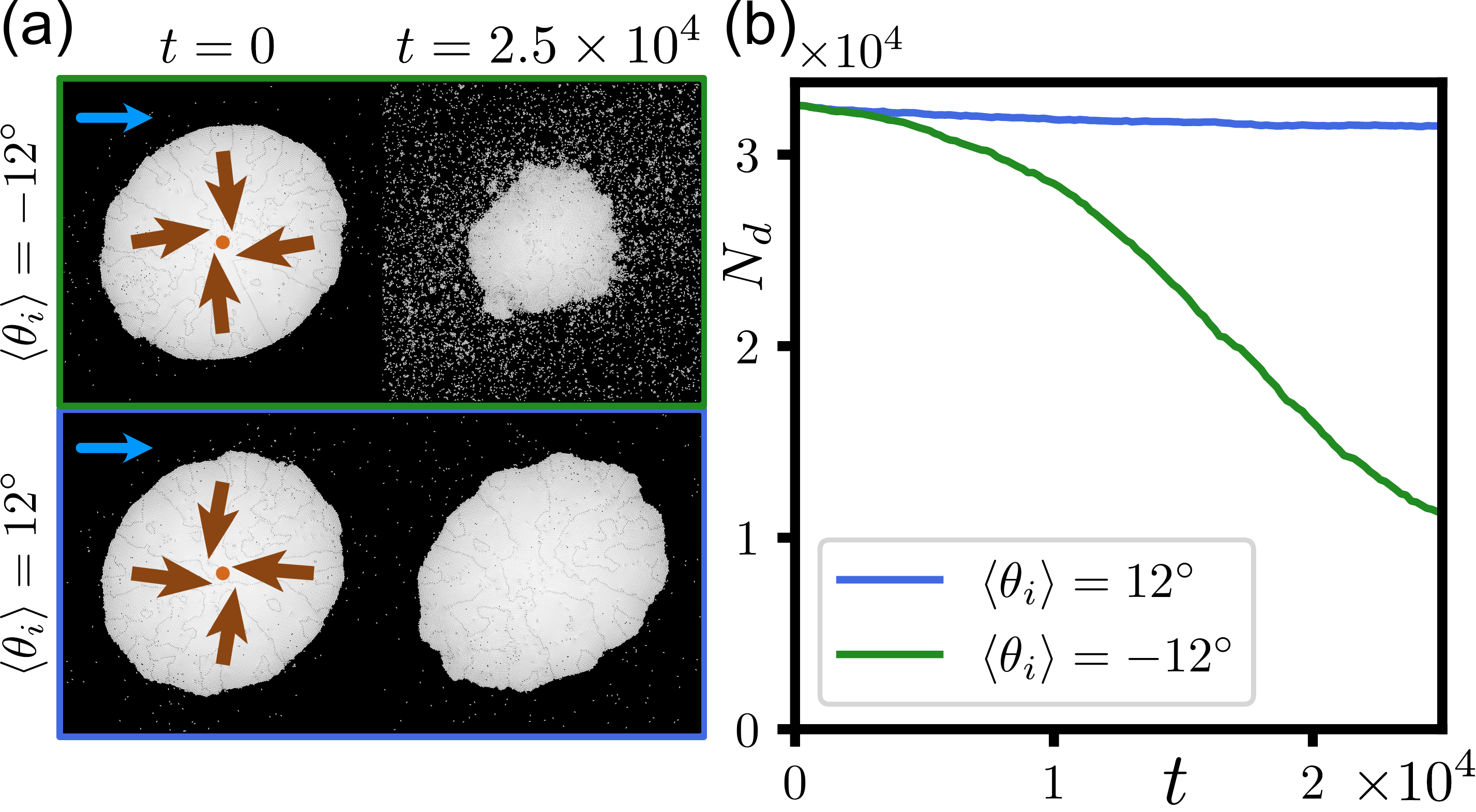}
\caption{
\textbf{Stability of clusters.}  (a) Snapshots illustrating the time evolution of a cluster of $N_d=32,768$ dumbbells prepared with initial  $\langle\theta_i\rangle=-12^\circ$ and   $\langle\theta_i\rangle=12^\circ$ (top and bottom row, respectively), at $t=0$ and $2.5 \times 10^4$ (first and second column), under shear rate $\textrm{Pe}_s=0.4$ and Pe=200.  Brown arrows illustrate the cluster polarization at $t=0$, while the red dot indicate the cluster center of mass. Finally, Blue arrows indicate the direction of shear.
Temporal evolution of the cluster mass ($N_d$) for initial negative (green) and positive (blue) $\langle\theta_i\rangle$ cases of (a).  }

\label{Fig5}
\end{figure}
Thus, chirality breaking, so far observed in the rotation of clusters aligning with the shear direction, manifests also in an unexpected way in the internal organization of dumbbells within the clusters, as indicated by a small non-zero $\langle\theta_i\rangle$.

The latter can be attributed to the way shear affects cluster stability: dumbbells oriented with positive $\langle\theta_i\rangle$ enable clusters to become  more resistant to shear, as they generate an opposite active torque which counteracts the global distortion caused by the shear.   Conversely, when the active torque reinforces the shear, it enhances deformation and facilitates cluster breakup. To better prove this point, we construct  two highly polarized cluster, respectively with $\langle\theta_i\rangle=-12^\circ,12^\circ$, and applied the same shear $\textrm{Pe}_s=0.4$ with Pe=200, ~\ref{Fig5}(a). We find that indeed the cluster with negative  $\langle\theta_i\rangle$ is destroyed much more easily by shear, as can be seen looking at the cluster mass as a function of time in Fig.~\ref{Fig5}(b).

{\bf Conclusions} In conclusion, in this work we have investigated how external driving influences the emergence of self organized structural order in an active matter system. 
More specifically, we have considered 
the phase separation process of active Brownian dumbbells under external driving.
By examining the time evolution of the average domain size curves $R(t)$ we have observed that cluster growth is arrested by shear. Clusters in the stationary regime become smaller with increasing shear and with decreasing Pe.
At the same time, the magnitude of local polarization in clusters becomes smaller. We have found that the application of shear breaks the rotational chirality of clusters, with dumbbells tending to organize on average in such a way that torque imparted by activity opposes the one imparted by shear, making clusters more stable. 
These observations can be useful to suggest new strategies to control and direct the collective behavior of self-propelled systems in complex environments. 
Future explorations could consider a more thorough characterization of the rheology of the system and the full $\textrm{Pe}_s$-$\textrm{Pe}$-$\phi$ phase diagram, a comparison with active disks, and including hydrodynamic interactions\cite{Negro2022}, which in colloids has been observed to suppress MIPS \cite{Negro2022,theersSRD,navarro2014,gompperROADMAP,thee2018}. Additionally, it would be interesting to explore the response to external driving in a three-dimensional setup\cite{CaporusooNegro2024,Wiese2023} .

\section{acknowledgements}
We thank CINECA award under the ISCRA initiative (IsCb1 AcT) and supported by MIUR via the projects PRIN 2020/PFCXPE, PRIN 2022/HNW5YL and Quantum Sensing and Modeling for One-Health (QuaSiModO).

\appendix

\section{Growth kinetics under shear for P\lowercase{e}=100, 150 and different packing fractions}
\label{app:growth}
\begin{figure}[ht!]
\centering
\includegraphics[width=0.45\columnwidth]{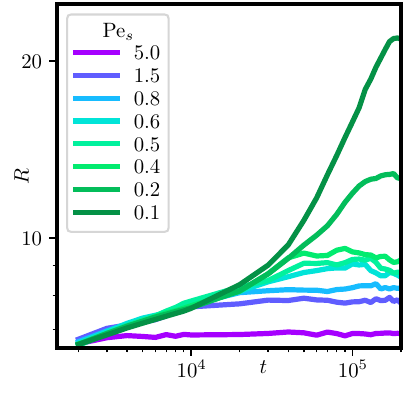}\hfill
\includegraphics[width=0.45\columnwidth]{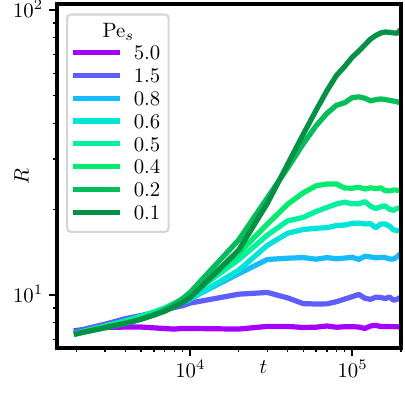}
\caption{Growth length $R(t)$ as a function of time for $\textrm{Pe}=100$ (left)
and $\textrm{Pe}=150$ (right). As in the $\textrm{Pe}=200$ case, $R(t)$
plateaus at late times to constant values $R_s$, which are affected by both
$\textrm{Pe}$ and $\textrm{Pe}_s$.}
\label{AppFig1}
\end{figure}

\begin{figure}[ht!]
\centering
\includegraphics[width=.5\columnwidth]{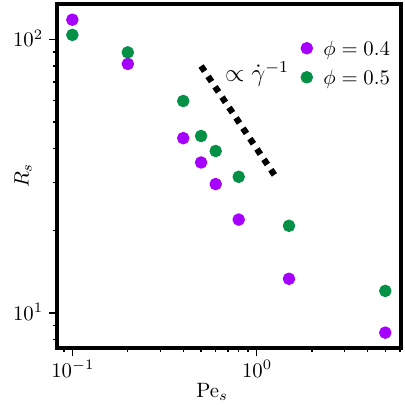}
\caption{Average value of cluster size $R_s$ as a function of shear rate for $\textrm{Pe}=200$,  and  global packing fractions $\phi = 0.4$ and $\phi = 0.5$. As dashed line, slope proportional to $1/\dot\gamma$.} 
\label{comparison_densities}
\end{figure}

Here we present additional results for the growth length $R(t)$, at
$\textrm{Pe}=100$ and $\textrm{Pe}=150$ (Fig.~\ref{AppFig1}). A behavior similar to the
$\textrm{Pe}=200$ case, shown in Fig.~2 of the main text, is observed:
$R(t)$ plateaus at late times to constant values $R_s$, which depend on both
$\textrm{Pe}$ and $\textrm{Pe}_s$.
To examine the effect of packing fraction, we vary it at fixed Péclet number ($\textrm{Pe}=200$). The results, shown in Fig.~\ref{comparison_densities}, indicate that in this regime the scaling discussed in the main text is still observed.

\section{Averages torques at different P\MakeLowercase{e} and P\MakeLowercase{e}$_s$}
\begin{figure*}[ht!]
\centering
       \includegraphics[width=0.6\columnwidth]{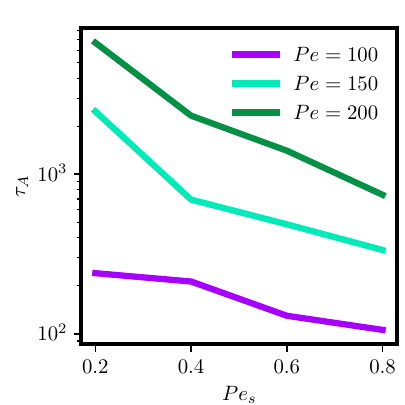}
       \includegraphics[width=0.6\columnwidth]{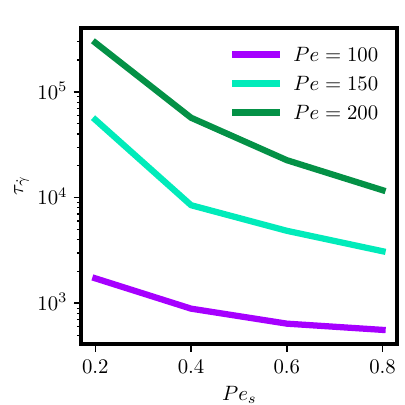}
       \includegraphics[width=0.6\columnwidth]{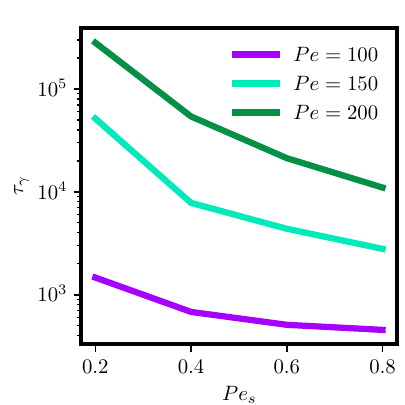}
\caption{Average values of ${\bm\tau}_A$, ${\bm\tau}_{\dot\gamma}$, ${\bm\tau}_\Gamma$ (left, middle, and right panels respectively) as a function of $\textrm{Pe}_s$, for different Pe in the key.}
\label{AppFigtorque}
\end{figure*}

Fig.~\ref{AppFigtorque} reports the average values of the three torque contributions, namely the active torque ${\bm\tau}A$, the shear–induced torque ${\bm\tau}{\dot\gamma}$, and the viscous drag torque ${\bm\tau}_\Gamma$ (shown in the left, middle, and right panels, respectively), as functions of the shear Péclet number $\textrm{Pe}s$. Results are presented for three different activity levels, $\textrm{Pe}=100,150,200$. In all cases, the qualitative behavior of the torques is the same.

\section{Theoretical formulas for torques}

We show how we estimate analytically the various torques, by approximating a cluster as a two-dimensional rigid body with elliptical shape.  The ellipse has  semi‑axes $r_a$ and $r_b$ and a particle density $\rho\sim0.9$ (corresponding to the binodal value of $\rho$ for all the values of Pe considered here~\cite{Petrelli2018}). Without loss of generality, we consider $r_a\ge r_b$. An infinitesimal mass element $dm=\rho dxdy$ of the ellipse is subjected to three different forces: an active force, a shear force and a drag force. Below we compute the total torque for each contribution, with respect to the ellipse center of mass, which is assumed to be the axis origin from now on.

We consider for each mass element an active force oriented with a uniform angle $\theta$ with respect to the center of the ellipse, and strength $F^a \sin\theta$ (see definition of $\theta_i$ in Fig.~\ref{Figscheme}).
The total torque is obtained integrating over the ellipse domain.  We consider a reference frame where the semi-axes $r_a$ and $r_b$ are aligned to the axis $x$ and $y$.  However, because the active force is independent on $x$ and $y$,  the integral remains invariant under a rotation of the ellipse   by an angle $\psi$ with respect to the $x$-axis (Fig.~\ref{Figscheme})). The integral is thus:

\begin{align}
    \bm\tau_A 
    &=  \int_{\frac{x^2}{r_a^2}+\frac{y^2}{r_b^2}<1}  \rho dxdy\,\,F^a\sin\theta\sqrt{x^2+y^2}=\\
    &= F^a\sin\theta \; r_a r_b\rho\int_{u^2+v^2<1} dudv\,\,\sqrt{r_a^2u^2+r_b^2v^2}=\\
    &= \frac{F^a\sin\theta}{3} \; r_a r_b\rho \int_0^{2\pi} d\phi \sqrt{ r_a^2\cos^2\phi + r_b^2\sin^2\phi}=\\
    &= \frac{F^a\sin\theta}{3} \; r_a^2 r_b\rho \int_0^{2\pi} d\phi\sqrt{1-\left(1-\frac{r_b^2}{r_a^2}\right)\sin^2\phi}=\\
    &= \frac{4F^a\sin\theta}{3} \; r_a^2 r_b\rho  E(e) \label{tau_active}
\end{align}
with $e=\sqrt{1-\frac{r_b^2}{r_a^2}}$  the eccentricity of the ellipse and $E(e)$ the complete elliptic integral of the second kind.

We next compute the torque exerted by a simple shear flow on the cluster ($\dot\gamma>0$). Here, the force acting on each mass element is $-y^2\Gamma\dot\gamma$, making the integral  dependent on the instantaneous orientation of the ellipse with respect to the $x$-axis.  We thus integrate using the domain depicted in Fig.~\ref{Figscheme}):

\begin{align}
    \bm\tau_{\dot\gamma} &= -\Gamma\,\dot{\gamma}\int_{\frac{(x\cos\psi-y\sin\psi)^2}{r_a^2}+\frac{(x\sin\psi+y\cos\psi)^2}{r_b^2}<1} \rho\,dx\,dy \,y^2=\\
    &=  -\Gamma\,\dot{\gamma}\rho\int_{\frac{u^2}{r_a^2}+\frac{v^2}{r_b^2}<1} \,du\,dv \,(-u\sin\psi+v\cos\psi)^2=\\
     &=-\Gamma\,\dot\gamma\,\rho\frac{\pi r_a r_b}{4}(r_a^2\sin^2\psi + r_b^2\cos^2\psi) \label{tau_gammadot}
\end{align}
We now compute the torque arising from viscous damping. Considering that the ellipse is rotating rigidly around its center with an angular velocity $\omega$, each mass element is subjected to a damping force $-\Gamma\omega (x^2+y^2)$, which is independent on the angle $\psi$.  Consequently, we can adopt the same frame as the one for the active torque.  

Accordingly, the total torque is:

\begin{align}
    \bm\tau_\Gamma &= -\Gamma\,\omega\,\int_{\frac{x^2}{r_a^2}+\frac{y^2}{r_b^2}<1}\,\rho \,dx\,dy\;(x^2+y^2)\\
    &=-\,\Gamma\rho\,\omega\,\frac{\pi r_a r_b}{4}(r_a^2+r_b^2). \label{tau_gamma}
\end{align}

\begin{figure*}[ht!]
\centering
       \includegraphics[width=0.6\columnwidth]{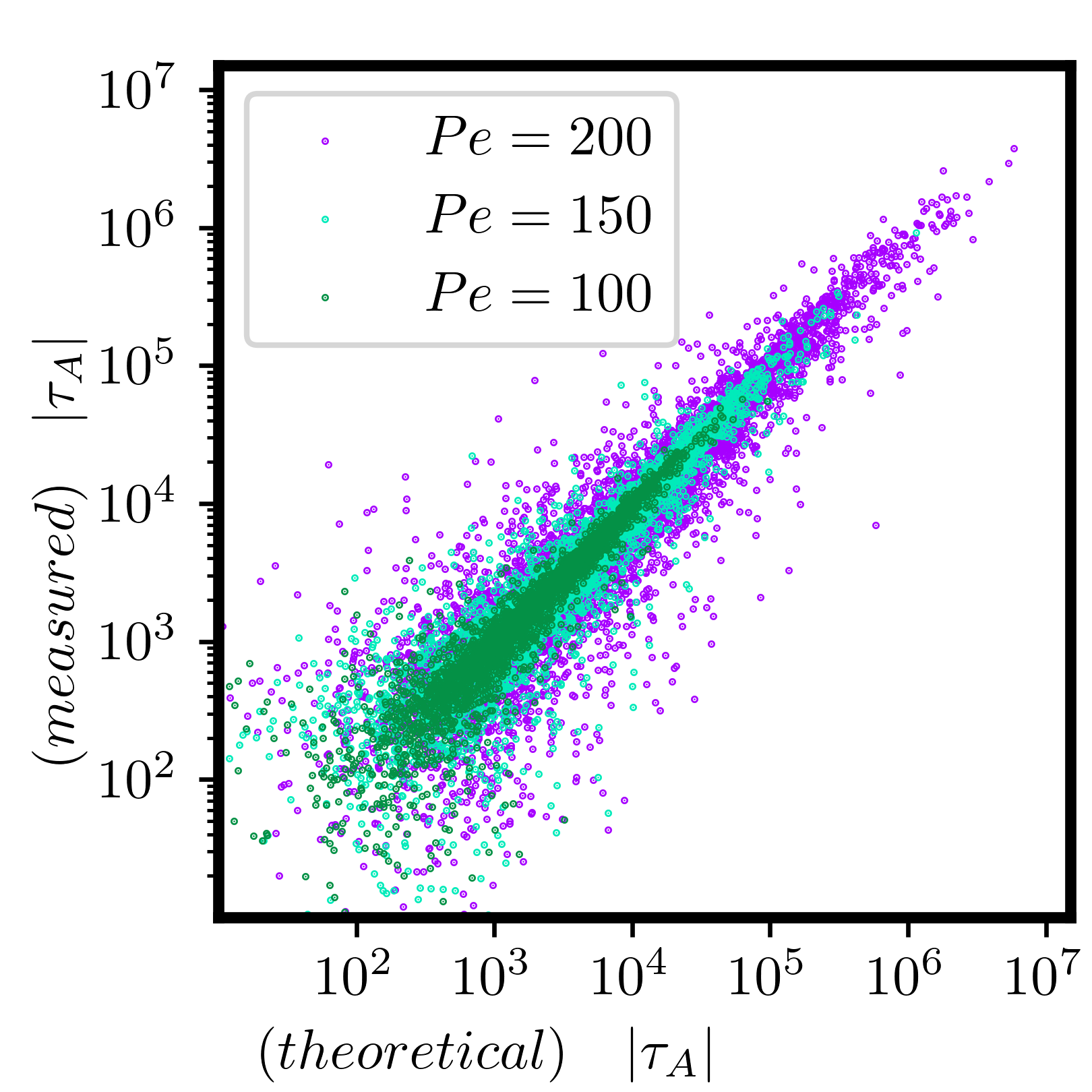}
       \includegraphics[width=0.6\columnwidth]{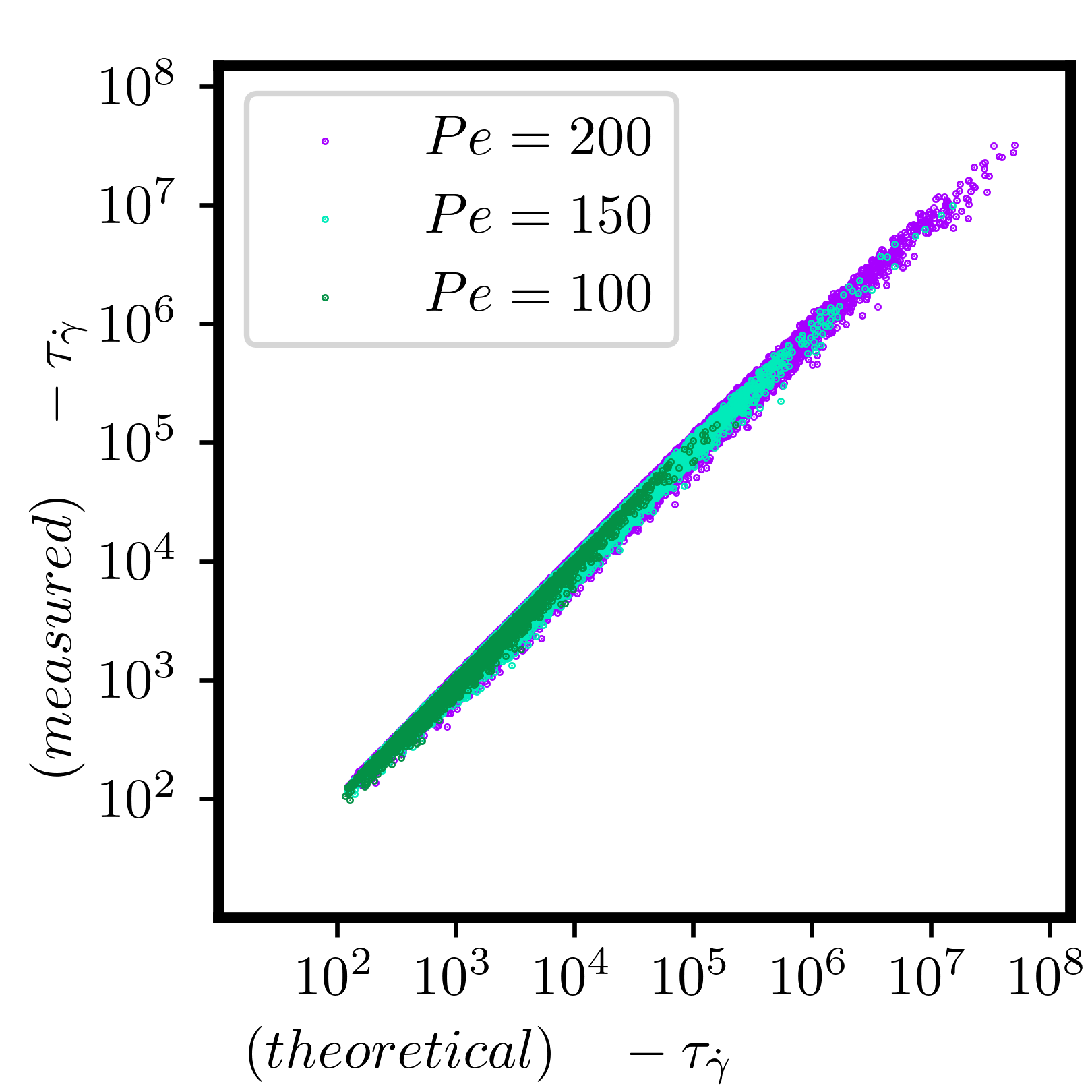}
       \includegraphics[width=0.6\columnwidth]{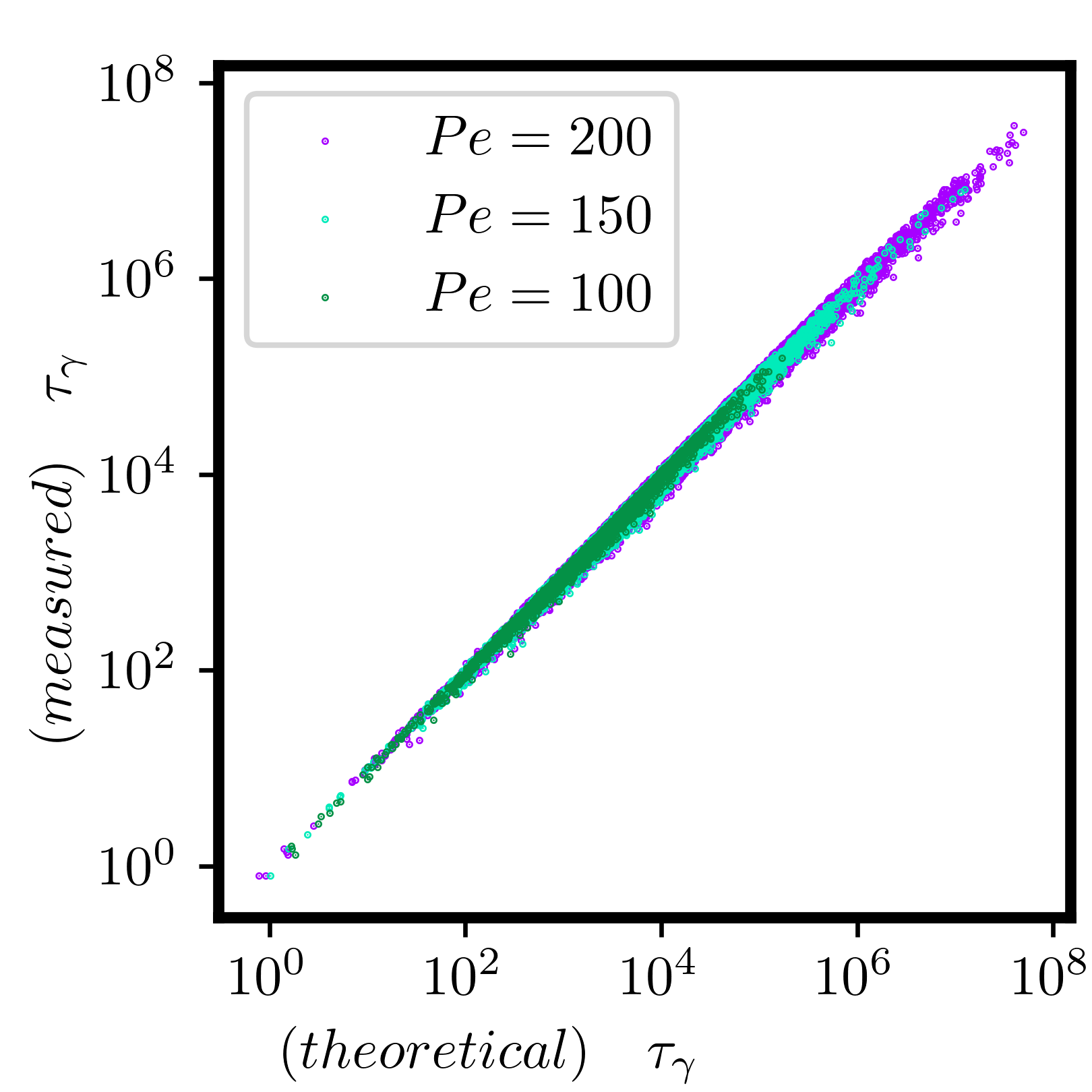}
\caption{Values of active torque (left), shear torque (middle) and drag torque (right) measured considering clusters  in the stationary regime, versus values estimated theoretically using the formulas \ref{tau_active},\ref{tau_gamma},\ref{tau_gammadot}}.
\label{Figpredizione}
\end{figure*}

Fig.~\ref{Figpredizione} compares the torques measured for individual clusters in the stationary regime against values estimated using formulas \ref{tau_active}, \ref{tau_gammadot}, and \ref{tau_gamma}. To obtain these estimates, for each cluster identified in the simulation, we directly substituted into the formulas the values of $r_a$, $r_b$ and $\psi$ derived from its inertial tensor. The average polarization angle $\langle\theta_i\rangle$, obtained averaging $\theta_i$ over all dumbbells within the cluster, was then used as the input for the angle $\theta$.

\begin{figure}[ht!]
\centering
       \includegraphics[width=0.6\columnwidth]{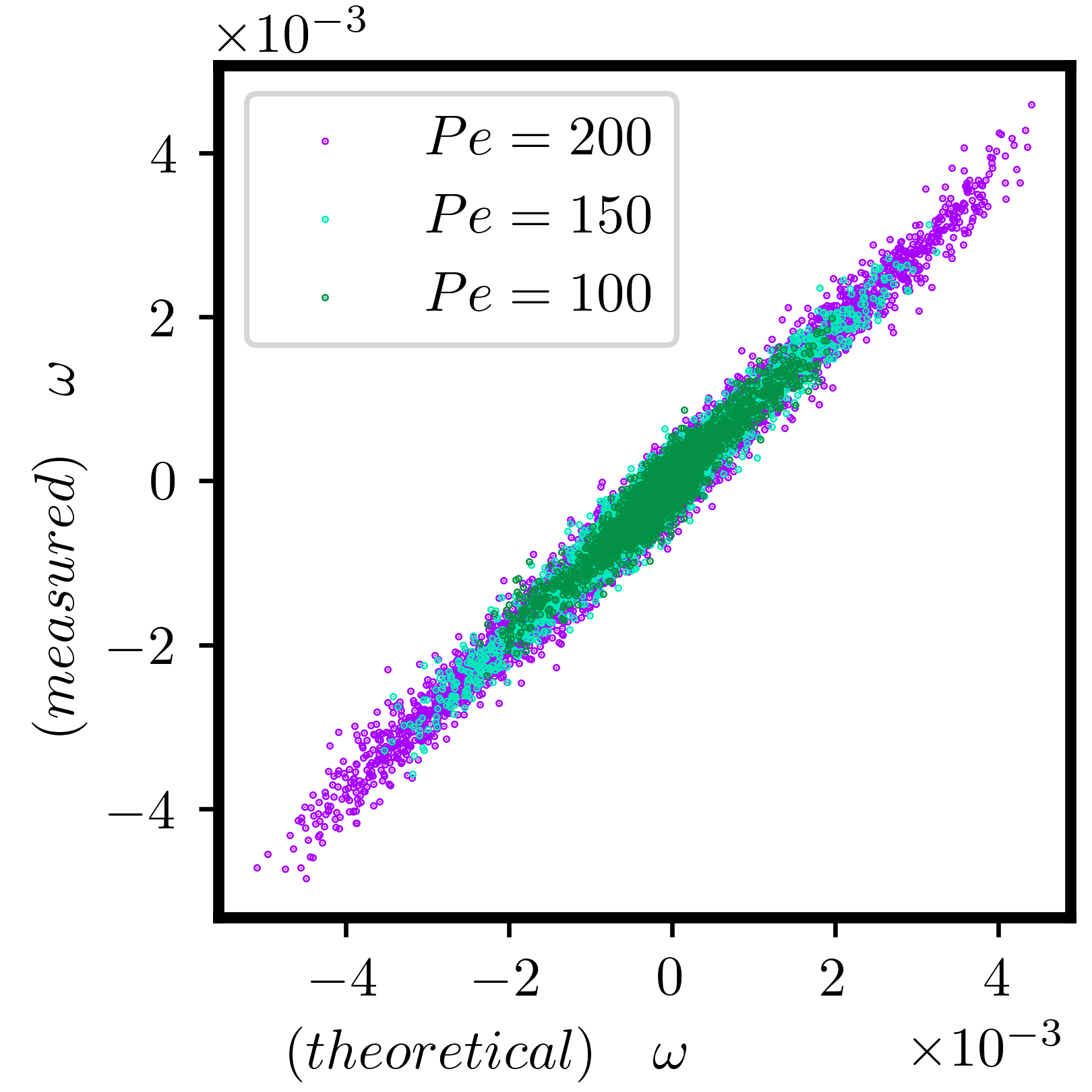}
\caption{Values of angular velocity of clusters measured for clusters in the stationary regime, versus values estimated theoretically using the formula \ref{prediction}\ , with $r_a$ and $r_b$ measured from simulations for each cluster at a given time. }
\label{Figpredizione1}
\end{figure}
In the overdamped regime, the three torques balance out $0=\tau_{\dot\gamma}+\tau_\Gamma + \tau_A$ and
we get  the angular velocity $\omega$ as a function of $\psi$:
\begin{equation}
    \omega=-\dot\gamma \frac{r_a^2\sin^2\psi + r_b^2\cos^2\psi}{r_a^2+r_b^2} + \frac{16}{3}\frac{F^a \sin\theta\, r_a E(e)}{\pi\Gamma(r_a^2+r_b^2)}.
    \label{prediction}
\end{equation}
Again, one can compare this formula with $\omega$ for clusters in the stationary regime, Fig.~\ref{Figpredizione1}, obtaining a good agreement.
Equation \ref{prediction} is  a first order differential equation of the form:
\begin{equation}
    \dot\psi=-A\sin^2\psi+B
\end{equation}
with $A=\dot\gamma \frac{r_a^2- r_b^2}{r_a^2+r_b^2} $ and $B=-\dot\gamma \frac{r_b^2}{r_a^2+r_b^2}+\frac{16}{3}\frac{F^a \sin\theta\, r_a E(e)}{\pi\Gamma(r_a^2+r_b^2)}$. Note that $A\ge0$, while $B$ can be in general either positive or negative. For typical parameters of our system, $A\approx 0.35\,\dot\gamma$ and  $B\approx-0.33$.

Such an equation admits a periodic solution of $\psi$ for $B<0$, with $\omega<0$. In this case  the cluster rotates clockwise since the shear torque is larger than the active torque. This is the typical case that we observe in our simulations. 
Similarly, for $B>A$, we would have a periodic solution of $\psi$ with $\omega>0$. In this case, the cluster rotates counter-clockwise, since the active torque prevails over the shear torque. 
For $0<B<A$, the equation admits a constant value of $\psi$, with the cluster remaining at a fixed $\psi$ value, and thus $\omega=0$.

In the case of an oscillating $\psi$,
corresponding to the typical case observed in simulations, one can easily obtain the average angular velocity over a period (both $B<0$ and $B>A$):
\begin{equation}
    \hat\omega=B\sqrt{1-\frac{A}{B}}  \ \rm{.}
    \label{aveomega}
\end{equation}

Finally, we expand these formula in the case of a small eccentricity $e$. This can be obtained transforming $r_a=r(1+\alpha)$ and $r_b=\frac{r}{1+\alpha}$, and noting that $e^2=1-\frac{b^2}{a^2}=1-\frac{1}{(1+\alpha)^4}$. Expanding~\ref{aveomega} with respect to $e$, we obtain:
\begin{equation}
    \hat\omega=\omega _{0}+e^{4}\left( -\dfrac{5}{48}\dfrac{F^a \sin \theta }{\Gamma r}-\dfrac{\dot\gamma^2}{32}\dfrac{1}{\omega _{0}}\right)+O(e^6) \rm{,}
\end{equation}
with $\omega _{0}=\frac{4}{3}\frac{F^a \sin{\theta}}{\Gamma r} - \frac{\dot\gamma}{2}$ the value of the angular velocity for a circle of radius $r$.

\bibliographystyle{apsrev4-1}
\bibliography{Refs_new.bib}

\begin{thebibliography}{75}
\bibitem [{not()}]{noterigid}%
  \BibitemOpen
  \href@noop {} {\ }\bibinfo {note} {In the case of reference [59], a MIE potential was used, and dumbbell’s connectivity was taken care of with the RATTLE scheme which makes the bond completely rigid, while here we use a WCA potential and extensive FENE potential for bonds.}\BibitemShut {Stop}%
\end{thebibliography}

\end{document}